# Zero Replacement and Latent Profiles in Short Media Diaries: Checking Recording Status Before User Segmentation

Eunjeong Song · Sehee Hong

Department of Education, Korea University, Seoul, Republic of Korea

Eunjeong Song, ORCID iD: https://orcid.org/0000-0002-2302-3227

Sehee Hong, ORCID iD: https://orcid.org/0000-0001-5468-8398

**Abstract**

Short media diaries contain many zeros because many activities go unrecorded during the observation period. Log-ratio analysis of time allocation requires these zeros to be replaced with small positive values; as a result, all respondents who did not record an activity receive nearly the same value. We examined how such zero replacement affects classification in latent profile analysis and propose checks to perform before profiles are interpreted. Three-day diaries of 8,089 adults from the 2025 Korea Media Panel Survey were converted into nine-part time-use compositions. Gaming was not recorded by 88.9% of the respondents, and in every model with two to eight profiles and a covariance matrix held equal across profiles, respondents who recorded gaming and those who did not were classified into different profiles. Even when the replacement value equaled the minimum recording unit (5 min/day), the two-profile classification largely matched gaming recording status, and the two groups separated more sharply as the replacement value decreased, until the separation was complete. Before naming profiles, researchers should therefore check the proportion of respondents who did not record each activity and its replacement value, the same proportion within each profile, and the distribution of usage time among those who recorded the activity.

*Keywords:* media user segmentation, media diary, time-use compositional data, latent profile analysis, zero replacement

**Zero Replacement and Latent Profiles in Short Media Diaries: Checking Recording Status Before User Segmentation**

Classifying users by the combinations of media and content they use is a common task in both media research and the content industry. The media repertoire perspective considers not only how much people use media but also which media they combine and how they divide their time among them (Hasebrink & Popp, 2006; Taneja et al., 2012). In Korea, studies have classified media repertoires (Shim, 2007), compared types of paid video-on-demand (VOD) users (Lee, 2020), analyzed patterns of content consumption on over-the-top (OTT) services (Lim et al., 2023), segmented OTT users (Chong et al., 2024), estimated latent classes of gaming-time trajectories (Hong & No, 2013), and identified types of social networking service (SNS) use among older adults in the 2025 Korea Media Panel Survey (Park, 2026). Unlike conventional cluster analysis, latent class and latent profile models compare solutions with different numbers of groups on statistical criteria and express each person's membership as a probability (DiStefano & Kamphaus, 2006; Vermunt & Magidson, 2002). Because the resulting types are used to select target groups for programming and advertising, the wrong audiences may be targeted if a type's name does not match its actual composition. We show that replacing the zeros in short media diaries can make latent profiles, and the names given to them, diverge from actual media use.

A media diary shows the time recorded within a fixed observation period, not long-term usage time. When the observation period is short, many activities have zeros. The same holds when users are segmented with platform logs (Chong et al., 2024), in which a service not used during the observation period is recorded as zero. Log-ratio analysis, which takes logarithms of ratios between activity times, cannot handle zeros, so the zeros are replaced with small positive values. Respondents who did not record an activity then take nearly the same value and can form a profile of their own (see the Effects of Zero Replacement on Profile Classification section). A researcher who looks only at profile means and names may then mistake differences in recording status for differences in the level of usage time.

## Time-Use Compositional Data and Zero Replacement

Because the 1,440 min of a day are divided among activities, more time spent on one activity means less time for the others. Data of this kind, in which the information lies in the relative sizes of the parts of a whole, are called *compositional data* (Aitchison, 1986). The *isometric log-ratio* (ilr) transformation maps a composition of $D$ parts with a fixed total onto $D - 1$ orthogonal coordinates (Egozcue et al., 2003). A *sequential binary partition* (SBP) first splits the parts into two groups and then repeatedly splits each group into two, and each ilr coordinate is the log-ratio of the geometric means of the two groups formed at one split, multiplied by a factor that depends on the group sizes (Egozcue & Pawlowsky-Glahn, 2005).

In this study, an activity is *recorded* by a respondent if it appears in the respondent's diary at least once during the 3-day observation period and *not recorded* if it never appears. We call respondents who recorded an activity *recorders* of that activity and the others *nonrecorders*. The zeros in time-use data are a mixture of zeros for activities that a person never does (*essential zeros*) and zeros for activities that the person does but did not record during the observation period (*count zeros*; Martín-Fernández et al., 2011). Because the diary is kept in 15-min slots, briefer use can also go unrecorded. It has long been recognized that the time recorded in a short diary differs from long-term usage time (Frazis & Stewart, 2012; Gershuny, 2012). Not recording an activity therefore does not mean not doing it.

Because log-ratios cannot be computed for zeros, the zeros are replaced with small positive values; this is called *zero replacement*, and the value inserted is called the *replacement value*. Two common methods are multiplicative replacement, which inserts a small value δ for each zero and shrinks the nonzero parts proportionally so that the total is preserved (Martín-Fernández et al., 2003), and geometric Bayesian-multiplicative (GBM) replacement, which sets the replacement value by using the share of the part in the rest of the sample as prior information (Martín-Fernández et al., 2015). In this study, multiplicative replacement inserts the same δ for every part. A replacement value is a computational device, not an estimate of the nonrecorders' usage time. When nearly the same replacement value is inserted for all nonrecorders of a part, they pile up at a single point on the log scale of that part; we call this concentration a *point mass*. We call the daily average time implied by a single 15-min record in 3 days (5 min) the *minimum recording unit* and the difference ln(minimum recording unit) − ln(replacement value) the *log gap*. The larger the log gap, the farther apart recorders and nonrecorders lie in ilr coordinates.

## Latent Profile Analysis and the Number of Profiles

Latent profile analysis (LPA) models the distribution of continuous indicator variables as a finite mixture of normal distributions and estimates profile-specific means and variances together with each person's posterior probabilities of profile membership (Masyn, 2013). Assigning each person to the profile with the largest posterior probability is called *modal assignment*. The number of profiles is chosen by weighing information criteria, likelihood ratio tests, profile sizes, and interpretability (Masyn, 2013; Nylund et al., 2007). Lower values of the information criteria—the Akaike information criterion (AIC), the Bayesian information criterion (BIC), and the sample-size-adjusted BIC (aBIC)—indicate better models (Nylund et al., 2007); when the criteria keep decreasing as profiles are added, the point at which the decrease levels off is used as a guide (Masyn, 2013, p. 571; Nylund-Gibson & Choi, 2018, p. 443). The likelihood ratio tests compare a $K$-profile model with a $(K - 1)$-profile model. They include the adjusted Lo–Mendell–Rubin likelihood ratio test (aLMR; Lo et al., 2001) and the parametric bootstrap likelihood ratio test (BLRT; McLachlan & Peel, 2000), and a significant result favors the $K$-profile model. In simulation studies, BIC among the information criteria and the BLRT among the tests identified the correct number of profiles most accurately (Nylund et al., 2007). If the smallest profile contains less than 5% of the sample—the lower end of the recommended minimum of 5%–8% (Nylund-Gibson & Choi, 2018, p. 447)—or fewer than 25 cases (Berlin et al., 2014), researchers examine whether adding it yields any substantive gain. Entropy, an index of classification quality, is not used to decide the number of profiles (Masyn, 2013, p. 569).

### Effects of Zero Replacement on Profile Classification

When normal mixture models approximate nonnormal distributions, they can extract profiles that do not exist (Bauer & Curran, 2003) or distort profile-specific estimates (Nam & Hong, 2021). Zero replacement produces markedly nonnormal distributions. The nonrecorders of a part form a point mass on its log scale, and when the replacement value is much smaller than the minimum recording unit, an empty interval also opens between them and the recorders. Respondents with zeros in several parts also share the same value on any coordinate built only from those parts. Mixture models are therefore expected to split off these isolated groups of nonrecorders before they capture differences in usage time. We use *separation* to mean that recorders and nonrecorders are classified into different profiles. Only a comparison with the original records can reveal what the profiles were split by.

**The Present Study**

The choice of zero-replacement method changes the distances between observations and the covariance structure (Lubbe et al., 2021; Rasmussen et al., 2020), and zero replacement before a log-ratio transformation has been reported to produce spurious clusters in cluster analysis (Tauber, 1999). However, few studies have varied both the replacement value and the number of profiles and checked, respondent by respondent, whether latent profile classification follows recording status for a part in which most values are zero. We applied the procedure of Gupta et al. (2020), which combines LPA with the ilr transformation widely used in time-use research (Dumuid et al., 2020), to short media diaries and addressed two research questions.

*Research Question 1:* Do latent profiles estimated from 3-day diaries divide respondents by whether they recorded an activity, regardless of the number of profiles?

*Research Question 2:* How does this separation by recording status change with the size of the replacement value?

By comparing individual classifications with the original diary records, we show that, after zero replacement in short diaries, latent profiles divide respondents by recording status rather than by the level of usage time at every number of profiles examined ($K$ = 2–8) and that most of this separation persists even when the replacement value equals the minimum recording unit (5 min/day). Because such classifications look very clear-cut according to a classification-quality index (entropy), we also propose a three-step checklist for comparing profiles with the original records.

# Method

## Data

We used the public-use data of the 2025 Korea Media Panel Survey (Han et al., 2025; Korea Information Society Development Institute [KISDI], n.d.). The survey's media diary divides each day into 96 time slots of 15 min (hereafter, slots) and records sleep and media activities in each slot for 3 days; one primary and one secondary activity can be recorded per slot. Of the 8,411 respondents who completed all 3 diary days, we analyzed the 8,089 adults aged 19 years or older (3,956 households). The days of the week covered by the diary differed across respondents, and 43.7% of the respondents had no weekend day among their diary days. The analysis sample was 55.2% female, with a mean age of 56.1 years (*SD* = 17.6); with cross-sectional survey weights, these figures were 50.6% and 51.1 years.

Survey weights were not used for model selection or classification, because the research questions concern classification within the sample and because the BLRT is not available for weighted models (Muthén & Muthén, 1998–2017, p. 819). Weights were applied only to population estimates such as profile shares and mean usage times, and confidence intervals accounted for clustering within households. As a secondary analysis of de-identified public-use data, the study was exempted from review by the Korea University Institutional Review Board (Exemption No. KUIRB-2026-0656-01).

## Time-Use Composition and Zero Replacement

An activity's daily usage time (hereafter, time) is the number of slots assigned to it over the 3 days multiplied by 15 min and divided by 3; the time spent gaming is called *gaming time*.

Because summing simultaneous activities would make a day longer than 1,440 min, only primary activities were used, and recording status is likewise based on primary activities. The time-use composition consists of nine parts: sleep, broadcast viewing, video/personal broadcasting, audio, communication, search/SNS/commerce, gaming, reading/learning/production, and other waking activities (Table 1; see Table A1 in Appendix A for the activity codes). The seven parts from broadcast viewing to reading/learning/production are called *media activities*. Other waking activities cover the waking time not spent on these seven activities and include minor media activities that were not kept as separate parts (4.3 min/day on average).

The main analysis applied GBM replacement (Martín-Fernández et al., 2015) to the slot counts (288 slots over 3 days); when a replacement value exceeded the smallest observed nonzero share of its part, it was capped at 65% of that share (Palarea-Albaladejo & Martín-Fernández, 2015). Appendix A gives the equations. The sensitivity analysis used multiplicative replacement (Martín-Fernández et al., 2003) with the same value δ for every zero, varying δ in nine steps from 5 to 0.1 min/day (log gap 0–3.9). The steps include the conventional value of 3.25 min/day, which treats the minimum recording unit as the detection limit and inserts 65% of it (Martín-Fernández et al., 2003). With both methods, the nonzero parts were rescaled to keep the daily total at 1,440 min, and the nine parts were then transformed into eight ilr coordinates with the SBP in Table A2.

**Table 1**

*Daily Time Allocation, Proportion of Nonrecorders, and Zero-Replacement Values*

| Part | *M* (*SD*), min/day | *Mdn*, min/day | Nonrecorders, % | GBM replacement, min/day | Log gap |
|---|---|---|---|---|---|
| Sleep | 493.8 (67.8) | 485.0 | 0.0 | | |
| Broadcast viewing | 194.9 (143.9) | 170.0 | 7.1 | 3.25 | 0.43 |
| Video/personal broadcasting | 41.7 (60.8) | 20.0 | 33.5 | 2.85 | 0.56 |
| Audio | 32.1 (73.6) | 0.0 | 56.6 | 2.20 | 0.82 |
| Communication | 49.0 (39.0) | 40.0 | 4.8 | 3.35 | 0.40 |
| Search/SNS/commerce | 23.1 (46.6) | 5.0 | 47.6 | 1.58 | 1.15 |
| Gaming | 6.3 (28.1) | 0.0 | 88.9 | 0.43 | 2.46 |
| Reading/learning/production | 65.8 (118.1) | 0.0 | 55.3 | 4.49 | 0.11 |
| Other waking activities | 533.4 (163.4) | 550.0 | 0.1 | 3.25 | 0.43 |

*Note.* $N = 8{,}089$. SNS = social networking service; GBM = geometric Bayesian-multiplicative. Daily minutes are 3-day totals divided by 3. Means, medians, and proportions of nonrecorders are unweighted and were computed from the diary data before replacement. The nonrecorders of a part are the respondents who never recorded it as a primary activity in the 3 days. The GBM replacement value is the geometric mean of the values assigned to zero cells; log gap = ln 5 − ln(replacement value), where 5 min/day is the minimum recording unit (one 15-min slot in 3 days). Sleep has no zeros, so its replacement value and log gap are not applicable.

## Latent Profile Analysis

The indicators were the eight ilr coordinates; we refer to a model with *k* profiles as the $K = k$ model. By default, Mplus constrains indicator variances to be equal across profiles in LPA and assumes *local independence*, that is, that the indicators are mutually independent within each profile (Muthén & Muthén, 1998–2017, p. 183). However, ilr coordinates built from different SBPs are orthogonal transformations of one another (Egozcue & Pawlowsky-Glahn, 2005; Egozcue et al., 2003), so under local independence the classification depends on which SBP is chosen (see Table B1 in Appendix B). Freely estimating all covariances among the indicators makes the classification invariant to the SBP, so the main analysis used a model in which this covariance matrix is held equal across profiles (hereafter, the *common-covariance model*; the class-invariant unrestricted specification in Masyn, 2013, p. 585). A model with profile-specific covariance matrices (hereafter, the *profile-specific covariance model*) had a smaller BIC, but only because it fitted the point mass with a very small variance in a profile made up solely of nonrecorders (Masyn, 2013, p. 590); we therefore report it only as a sensitivity analysis (Table B1).

The models were estimated by robust maximum likelihood (MLR) in Mplus 9.1 (Muthén & Muthén, 1998–2017) with 3,000 random sets of starting values and 750 final-stage optimizations, and we checked that the best log-likelihood value was replicated (Masyn, 2013, pp. 563–565); Appendix C reports the number of replications. For the final model, the same best log-likelihood was replicated when the number of random starts was doubled. Class enumeration was performed without covariates (Jang & Hong, 2023) by comparing the $K = 1$–8 models on the criteria described in the Latent Profile Analysis and the Number of Profiles section. The aLMR and BLRT were obtained with the TECH11 and TECH14 options of Mplus (Asparouhov & Muthén, 2012); because of the computational burden, the BLRT was computed only for $K = 2$–6, with 200 bootstrap draws. Profile shares were computed by applying survey weights to the modal assignments; only the smallest profile proportion in Table 2 is a model-based estimate. Profile names state recording status first, and recorder profiles are further labeled by their relative level of gaming time within the 3-day observation period (long, medium, or short).

**Recording-Status Check**

We checked in three steps whether the profiles were divided by recording status (hereafter, the *recording-status check*):

1. For each part, examine the proportion of nonrecorders, the replacement value, the log gap, and the median time among recorders.
2. Compare individual classifications with the original records to determine whether each profile consists of recorders or of nonrecorders.
3. Within profiles of recorders, examine the distribution of time before replacement.

We applied Step 2 to gaming, the part with the most nonrecorders, in the $K$ = 2–8 models; for the $K$ = 5 model, we also computed the proportion of nonrecorders by profile for all seven media activities.

A profile consisting only of recorders is called a *recorder profile*, one consisting only of nonrecorders a *nonrecorder profile*, and one containing both a *mixed profile*; separation is *complete* when there is no mixed profile. Agreement between a classification and recording status was measured with the adjusted Rand index (ARI; Hubert & Arabie, 1985), which equals 1 for perfect agreement. In the sensitivity analyses, the classifications from models that differed in survey weighting, zero-replacement method, or covariance structure were compared with that of the main analysis (Table B1). In the sensitivity analysis that varied the replacement value over nine steps (Research Question 2), we estimated, with the same settings as the main analysis, a $K$ = 2 model for comparison with the binary recording status and a $K$ = 5 model for comparison with the main analysis (Tables B2–B4).

## Transparency and Openness

The raw data can be downloaded from the KISDI Media Statistics Portal after agreeing to its download terms (KISDI, n.d.). Data, materials, and code are available in Harvard Dataverse (https://doi.org/10.7910/DVN/YIJEVR). During preparation of this work, the authors used DeepL to translate the initial draft and improve English readability, and OpenAI Codex to assist with refinement and error checking of the research code. The authors reviewed and edited all outputs and take full responsibility for the final manuscript and research code.

# Results

## Zeros and Point Masses

Of the respondents, 97.2% had a zero in at least one of the nine parts, and the mean number of zeros per respondent was 2.94. The proportion of nonrecorders was highest for gaming (88.9%), followed by audio (56.6%), reading/learning/production (55.3%), and search/SNS/commerce (47.6%; Table 1). Because GBM replacement values are nearly proportional to the overall share of a part (Appendix A), the replacement value equaled 6.8% of the part's mean time for each of the six media parts to which the cap did not apply; consequently, parts with shorter mean times had larger log gaps. The replacement value for gaming was 0.43 min/day and practically constant across respondents, so the 7,194 gaming nonrecorders formed a point mass on the log scale of gaming. The log gap for gaming was 2.46, or 2.6 times the standard deviation of log gaming time among recorders (0.958). The log gaps of the other media activities ranged from 0.11 to 1.15, whereas under multiplicative replacement ($\delta = 3.25$ min/day) every part had a log gap of 0.43. The point mass also appeared in the ilr coordinates: Because the ratio of the two replacement values does not depend on the respondent, the 4,144 respondents (51.2%) who recorded neither gaming nor reading/learning/production had exactly the same value on the coordinate contrasting these two parts ($z_8$; Appendix A).

## Number of Profiles

Across the *K* = 1–8 models (Table 2), the information criteria continued to decrease through *K* = 8, but each added profile reduced the BIC by 3,556–17,708 through *K* = 5 and by only 2,008–2,341 from *K* = 6 onward (see Figure C1 in Appendix C). The aLMR was significant through *K* = 6 but not from *K* = 7 onward, and the BLRT was significant for every *K* for which it was computed (2–6). The profile added at *K* = 6, however, consisted of 42 nonrecorders (model-estimated proportion 0.5%) and offered no interpretive gain. The smallest profile at *K* = 5 (2.5%) was also below 5%, but its 201 members had a mean posterior probability of .983 and could be interpreted as respondents who recorded long gaming times, so we selected *K* = 5 as the final model. The entropy of the *K* = 5 model was .994, high enough for modal assignment to be used in describing the profiles on variables not included in the model (time before replacement, age, and sex; Clark & Muthén, 2009).

**Table 2**

*Model Comparison by Number of Profiles*

| *K* | LL | Parameters | AIC | BIC | aBIC | aLMR *p* | BLRT *p* | Entropy | Smallest profile, % |
|---|---|---|---|---|---|---|---|---|---|
| 1 | −93,352.9 | 44 | 186,793.9 | 187,101.8 | 186,962.0 | | | | |
| 2 | −84,458.2 | 53 | 169,022.5 | 169,393.4 | 169,225.0 | < .001 | < .005 | 1.000 | 11.1 |
| 3 | −81,286.4 | 62 | 162,696.7 | 163,130.6 | 162,933.6 | .001 | < .005 | .996 | 5.3 |
| 4 | −79,468.1 | 71 | 159,078.2 | 159,575.1 | 159,349.4 | < .001 | < .005 | .997 | 2.4 |
| 5 | −77,532.8 | 80 | 155,225.7 | 155,785.6 | 155,531.3 | < .001 | < .005 | .994 | 2.5 |
| 6 | −76,322.0 | 89 | 152,822.1 | 153,444.9 | 153,162.1 | .002 | < .005 | .995 | 0.5 |
| 7 | −75,265.1 | 98 | 150,726.3 | 151,412.1 | 151,100.7 | .319 | — | .994 | 0.6 |
| 8 | −74,220.6 | 107 | 148,655.2 | 149,404.0 | 149,063.9 | .575 | — | .968 | 0.5 |

*Note.* All models are unweighted and use GBM replacement; the final model is $K = 5$. LL = log-likelihood; AIC = Akaike information criterion; BIC = Bayesian information criterion; aBIC = sample-size-adjusted BIC; aLMR = adjusted Lo–Mendell–Rubin likelihood ratio test; BLRT = parametric bootstrap likelihood ratio test; GBM = geometric Bayesian-multiplicative. The BLRT was computed for $K = 2$–6 with 200 bootstrap draws, so $p < .005$ is the smallest reportable value; dashes indicate that it was not computed. Smallest profile is the smallest model-estimated profile proportion. Blank cells are not applicable to the one-profile model.

## Profiles and Gaming Recording Status (Research Question 1)

At $K = 2$, the 895 gaming recorders and the 7,194 nonrecorders each formed a profile (Figure 1). At $K = 3$ and $K = 4$, the nonrecorders remained in a single profile, and only the recorders split further, into two and three profiles differing in gaming time, respectively; at $K = 5$, the nonrecorders also split, mainly by whether they recorded broadcast viewing, into groups of 441 and 6,753. At $K = 6$–8, small nonrecorder profiles of 42–45 respondents were added, and both recorders and nonrecorders were subdivided further, but no mixed profile appeared at any *K* examined (2–8; see Table C1). In the common-covariance model, then, complete separation by gaming recording status occurred regardless of the number of profiles; the number of profiles determined only how finely the recorders and the nonrecorders were each subdivided. The choice of $K = 5$ therefore does not affect this conclusion. In the profile-specific covariance models ($K = 2$ and 3), one nonrecorder profile appeared, but the remaining nonrecorders formed mixed profiles with recorders (Table B1).

**Figure 1**

*Profile Assignment and Gaming Recording Status in the Models With K = 2–5*

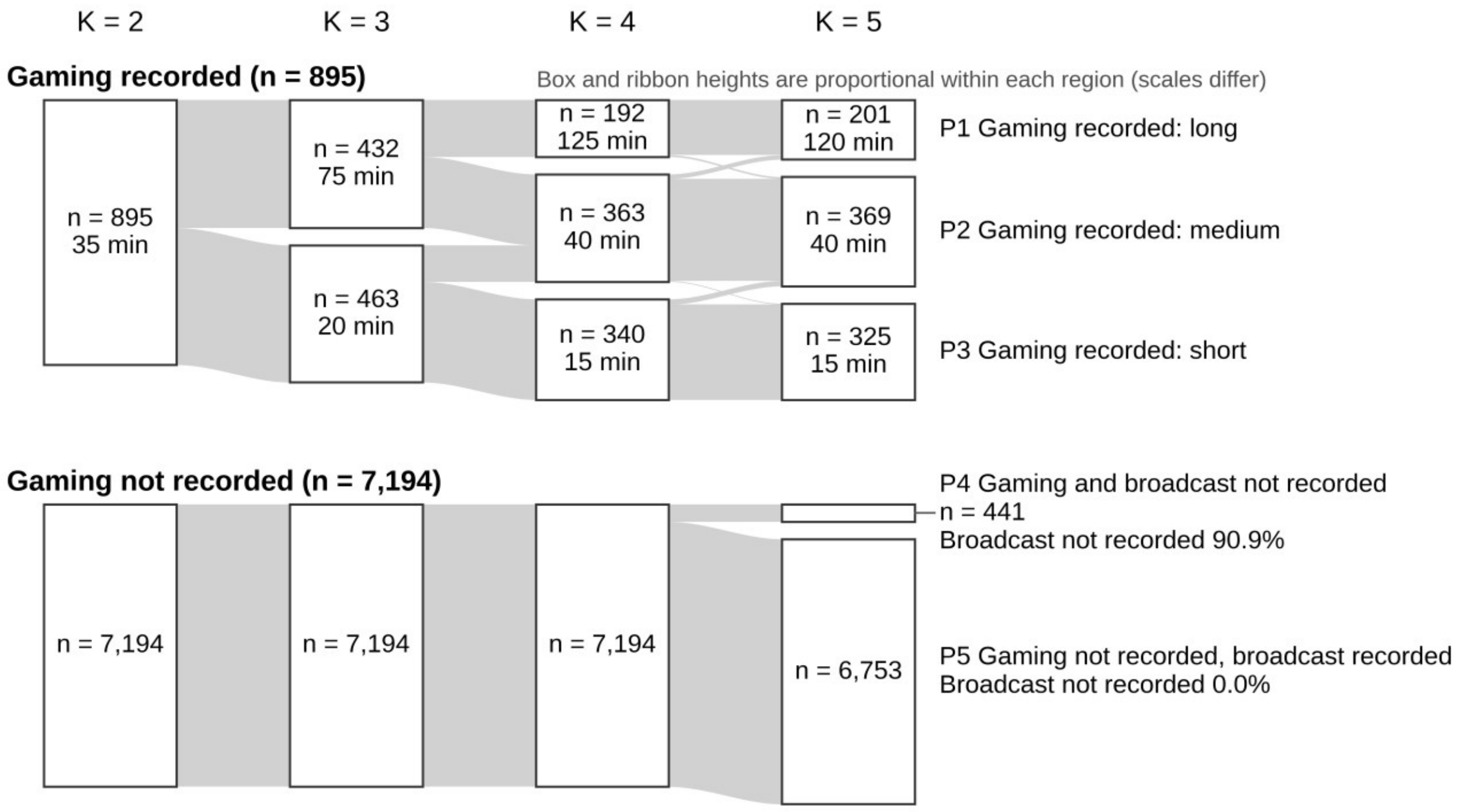

*Note.* Boxes represent profiles, and gray ribbons show how respondents moved between separately estimated adjacent models. Box and ribbon heights are proportional to the number of respondents within each region (gaming recorders, top; nonrecorders, bottom), and the two regions are drawn on different scales. Times in the boxes are median daily gaming minutes before replacement. Profile numbers and names at $K = 5$ follow Table 3.

## Usage Time and Recording Status by Profile

The mean posterior probabilities in the $K = 5$ model ranged from .983 to .998 (Table 3), and the survey-weighted model assigned 99.8% of the respondents to the same profiles (Table B1). Gaming recorders (P1–P3) made up 14.3% of adults (survey-weighted; 11.1% unweighted). Their gaming time ranged from 5 to 585 min/day; with survey weights, 23.5% recorded 15 min/day or less and 31.1% recorded 60 min/day or more. The three recorder profiles differed in their levels of gaming time, but their ranges overlapped at the boundaries (P1, 55–585 min/day; P2, 20–90 min/day; P3, 5–25 min/day; see Figure D1 in Appendix D), because classification is determined by the allocation of time across all nine parts rather than by gaming time alone.

**Table 3**

*Size, Posterior Probability, Gaming Time, and Proportion of Nonrecorders by Profile in the Five-Profile Model*

| Characteristic | P1 | P2 | P3 | P4 | P5 | Total |
|---|---|---|---|---|---|---|
| *n* | 201 | 369 | 325 | 441 | 6,753 | 8,089 |
| Weighted share, % | 3.0 | 5.8 | 5.4 | 6.4 | 79.4 | 100.0 |
| Mean posterior probability | .983 | .985 | .993 | .986 | .998 | |
| Gaming time, *Mdn* (IQR), min/day | 120 (95–160) | 40 (30–55) | 15 (10–20) | | | 35 (20–70) |
| Nonrecorders, % | | | | | | |
| Gaming | 0.0 | 0.0 | 0.0 | 100.0 | 100.0 | 88.9 |
| Broadcast viewing | 51.2 | 15.7 | 4.0 | 90.9 | 0.0 | 7.1 |
| Video/personal broadcasting | 14.4 | 14.1 | 15.4 | 12.9 | 37.3 | 33.5 |
| Audio | 42.8 | 33.9 | 38.5 | 42.0 | 60.1 | 56.6 |
| Communication | 13.9 | 3.8 | 1.8 | 13.6 | 4.1 | 4.8 |
| Search/SNS/commerce | 40.8 | 24.7 | 23.7 | 36.1 | 51.0 | 47.6 |
| Reading/learning/production | 39.3 | 35.0 | 36.3 | 29.9 | 59.4 | 55.3 |

*Note.* P1 = gaming recorded: long; P2 = gaming recorded: medium; P3 = gaming recorded: short; P4 = gaming and broadcast not recorded; P5 = gaming not recorded, broadcast recorded. Profiles are numbered by descending proportion of gaming recorders, then by descending mean gaming time, and then by descending total media time. Sizes (*n*) are unweighted counts under modal assignment, and weighted shares are the survey-weighted proportions of those assignments (95% confidence intervals in Table D1); shares are rounded separately (P1–P3 combined: 14.3%). Mean posterior probability is the posterior probability of a profile averaged over the respondents assigned to it. Gaming time is the median (interquartile range [IQR]) of daily gaming minutes before replacement among gaming recorders; the Total column covers all 895 gaming recorders. Gaming times and proportions of nonrecorders are unweighted. Blank cells are not applicable. SNS = social networking service.

All P4 members were gaming nonrecorders, and 90.9% of them (401 of 441) had not recorded broadcast viewing either. All 6,753 P5 members recorded broadcast viewing, so every one of the 401 respondents who recorded neither gaming nor broadcast viewing belonged to P4. We therefore labeled P4 “gaming and broadcast not recorded” and P5 “gaming not recorded, broadcast recorded.” In terms of usage time, P4 had the longest times of the five profiles for reading/learning/production (171.5 min/day) and video/personal broadcasting (134.7 min/day), whereas in P5 broadcast viewing (201.2 min/day) accounted for 48.7% of media time (survey-weighted means; Figure D2). Table D1 gives the survey-weighted profile shares with 95% confidence intervals and the mean age and percentage of men in each profile.

## Change in Classification With the Replacement Value (Research Question 2)

Even when the replacement value equaled the minimum recording unit, the $K = 2$ classification largely matched gaming recording status. As $\delta$ decreases from 5 to 0.1 min/day, the log gap for gaming increases from 0 to 3.9 (Figure 2). Under every condition, gaming was the media activity whose recording status agreed best with the $K = 2$ classification. When the log gap was 0 ($\delta = 5$ min/day), the ARI was .83, and 206 recorders (23.0%) were classified into the mixed profile, which consisted mostly of nonrecorders; all 206 had recorded 20 min/day or less. The ARI increased with the log gap and reached 1.00 under the GBM condition and for $\delta \leq 0.25$ min/day (Table B2).

At $K = 5$, the solutions for $\delta$ = 1–4 min/day consisted, like the main solution, of recorder profiles at three levels of gaming time and two profiles made up mostly of nonrecorders, and they agreed closely with the main solution (ARI = .97–.98; Table B3); at the conventional value of $\delta = 3.25$ min/day, 99.2% of the respondents were assigned to the same profiles as in the main solution (Table B1). For $\delta \leq 0.5$ min/day, the log gaps of the other parts grew as well, so the recording status of broadcast viewing and then that of video/personal broadcasting began to split the profiles; for $\delta \leq 0.25$ min/day, the distinction among levels of gaming time disappeared (Tables B3 and B4).

**Figure 2**

*Agreement Between the K = 2 Classification and Gaming Recording Status by Zero-Replacement Value δ*

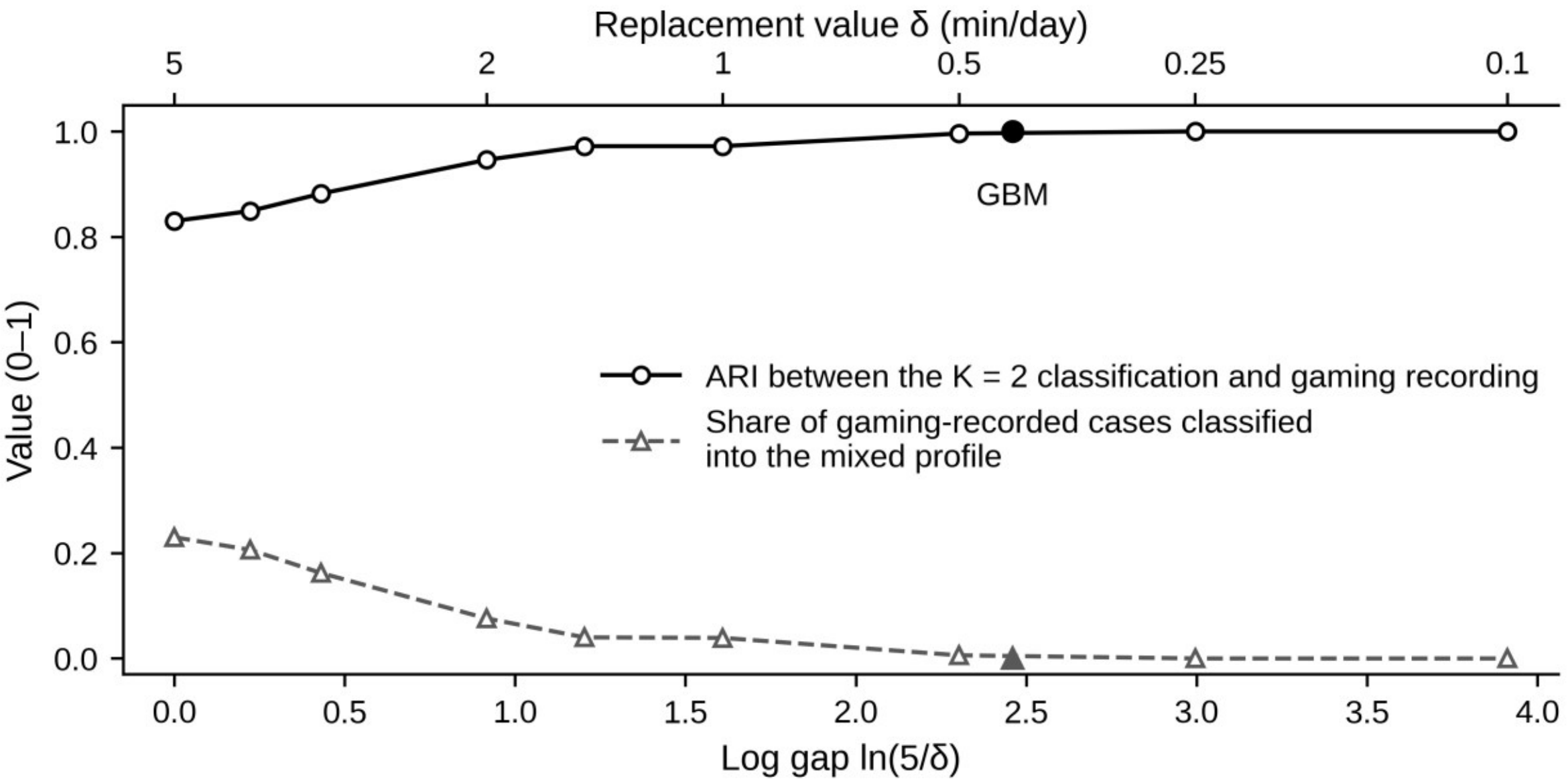

*Note.* Unweighted $K = 2$ models were estimated for each replacement condition. The bottom axis shows the log gap ln(5/δ), and the top axis shows δ (min/day). Filled markers indicate the GBM condition of the main analysis (gaming replacement value 0.43 min/day); lines connect only the multiplicative-replacement conditions. The dashed series is the proportion of the 895 gaming recorders classified into the mixed profile, which consists mostly of nonrecorders.

ARI = adjusted Rand index; GBM = geometric Bayesian-multiplicative.

## Discussion

### Main Findings and Implications for Content User Segmentation

Inserting nearly the same replacement value for all nonrecorders created a point mass in the log-ratio coordinates, and the common-covariance model split off this point mass first. Four results stand out. First, gaming recorders and nonrecorders were completely separated at every number of profiles examined (*K* = 2–8). Second, even when the replacement value equaled the minimum recording unit, so that there was no empty interval, the *K* = 2 classification largely matched gaming recording status (ARI = .83), and all recorders classified into the profile made up mostly of nonrecorders had recorded 20 min/day or less. Separation thus arose from the point mass itself rather than from an empty interval and became more complete as the log gap grew. Third, GBM replacement, which is widely used for count zeros, inserts values nearly proportional to the overall share of a part; it therefore inserted the smallest value (0.43 min/day) for gaming, the part with the shortest mean time, and widened its log gap (2.46). Fourth, judged by entropy (.994–1.000 at *K* = 2–7) and mean posterior probabilities (.983–.998 at *K* = 5), these solutions nonetheless appear to classify respondents very clearly. High entropy indicates that respondents are classified with little uncertainty, not that the profiles are valid, so the problem becomes visible only when the profiles are compared with the original records. Previous studies have shown that zero replacement changes the distances between observations and the covariance structure (Lubbe et al., 2021; Rasmussen et al., 2020) and can create spurious clusters (Tauber, 1999). We confirmed at the individual level that in normal-mixture LPA this effect appears as separation by recording status and that the size of the replacement value affects how complete the separation is.

These results show why profile names should not be read as attributes of users. Calling the recorder profiles “gamers” and the nonrecorder profiles “nongamers” overlooks two facts. First, the recorder profiles include respondents who recorded only 15 min/day or less (23.5%, survey-weighted). Second, the nonrecorder profiles (85.7%, survey-weighted) include occasional players who simply did not play during those 3 days (the count zeros described in the Time-Use Compositional Data and Zero Replacement section). Even among the gaming nonrecorders, 7.6% (549) answered “yes” (unweighted) to a person-level survey item that asked whether the respondent had used online game content in the past year, together with average monthly spending on it (Han et al., 2025, p. 191). Because the item appears to have been interpreted narrowly—only 39.4% of the gaming recorders answered “yes”—the nonrecorders probably include more players than this figure suggests. Targeting game advertising at the recorder profiles, for example, would miss such occasional players. For activities with many zeros or with replacement values far from the recorded values (gaming and broadcast viewing in these data), recording status can easily become the basis of segmentation, so if profiles formed in this way are used directly as target groups for programming or advertising, the wrong audiences may be targeted.

Table 4 summarizes what to check when users are segmented with short diary or log data. Steps 1–3 constitute the recording-status check described in the Method section, and the code available in Harvard Dataverse (see the Transparency and Openness section) contains functions that compute them. If the profiles divide respondents by recording status, researchers can respond in one of the ways listed as Step 4: They can state recording status in the profile names so that the profiles are not interpreted beyond that distinction; use a two-part factor mixture model, which models both whether an activity was recorded and how much time recorders spent on it (Kim & Muthén, 2009); or use a mixed-scale latent class cluster model with both categorical and continuous indicators (Vermunt & Magidson, 2002). Merging a part with many zeros into another part (Dumuid et al., 2020) makes the time levels of that activity indistinguishable, so it is suitable only for parts that are not the focus of segmentation.

**Table 4**

*Recording-Status Checklist for User Segmentation With Short Diary or Log Data*

| Step | Check | Implication | This study |
|---|---|---|---|
| 1. Parts | Proportion of nonrecorders, replacement value, log gap, and median time among recorders, for each part | A part with many nonrecorders, or with a replacement value far from the recorded values, can split the profiles by itself | Gaming: 88.9% nonrecorders, replacement value 0.43 min/day (log gap 2.46); broadcast viewing: 7.1% nonrecorders, replacement value 3.25 min/day vs. a median of 180 min/day among recorders |
| 2. Profiles | Proportion of nonrecorders of each part, by profile | Proportions at or near 0% or 100% indicate that the profiles divide respondents by recording status | Gaming: 0% or 100% in every profile for $K$ = 2–8; broadcast viewing ($K$ = 5): 0% in P5 and 90.9% in P4 |
| 3. Recorders | Distribution of time among recorders in each profile | Name profiles by both recording status and level of time | Ranges of 5–25, 20–90, and 55–585 min/day overlap at the boundaries |
| 4. Response | | State recording status in profile names, use a two-part or mixed-scale mixture model, or merge parts that are not the focus | Profile names state recording status |

*Note.* Steps 1–3 constitute the recording-status check described in the Method section; Step 4 lists possible responses to its results. Blank cells are not applicable.

## Limitations and Future Research

This study has three limitations. First, it focused on one 3-day diary data set and one part (gaming). Whether the same results hold for other parts with many zeros or with replacement values far from the recorded values, and for longer diaries or platform logs, remains to be examined. Respondents whose diary days included two weekend days had a somewhat higher gaming recording rate (12.2%, unweighted) than the other respondents (10.6%), so recording status may also depend on which days of the week are covered. Second, the results depend on the zero-replacement method and the covariance structure. We examined only GBM and multiplicative replacement ($\delta$ = 0.1–5 min/day), and GBM replacement treats the counts of consecutive slots as if they were independent trials, so zero-handling methods suited to diary data need further study. Log-ratio expectation-maximization (EM) replacement, which inserts a different replacement value for each person (Rasmussen et al., 2020), removes the point mass but still concentrates the nonrecorders below the minimum recording unit, so the same check is needed for such methods. Third, to keep the daily total fixed, we used only primary activities, which left out media time recorded as secondary activities (46.0 min/day on average). In addition, because the public-use data lack enumeration-district and stratum identifiers, the original sampling design could not be fully taken into account.

## Conclusion

When nearly the same small replacement value was inserted for the zeros of short diaries, a latent profile model with a covariance matrix held equal across profiles divided respondents by recording status rather than by the level of usage time at every number of profiles examined ($K = 2–8$), and most of this separation remained even when the replacement value equaled the minimum recording unit. Within the range examined (GBM and multiplicative replacement with $\delta = 0.1–5$ min/day), smaller replacement values made the separation more complete, and very small values let the recording status of other activities split the profiles as well. When users are segmented with short diary or log data, the results of the checks in Table 4 should therefore be reported along with profile means and names, so that recording status during the observation period is not mistaken for a long-term attribute of users.

# Appendix A

## Time-Use Composition, Coordinates, and Zero Replacement

Table A1 lists the media diary activity codes that make up each part of the time-use composition, and Table A2 gives the sequential binary partition (SBP) used to construct the eight isometric log-ratio (ilr) coordinates. The last section of this appendix gives the zero-replacement equations.

**Table A1**

*Composition Parts and Media Diary Activity Codes for Primary Activities*

| Part | Activity codes |
|---|---|
| Sleep | Sleep flag |
| Broadcast viewing | 1, 2, 3, 4, 5, 29, 37, 38 |
| Video/personal broadcasting | 9, 10, 41, 42 |
| Audio | 6, 7, 11 |
| Communication | 17, 18, 19, 43, 44 |
| Search/SNS/commerce | 20, 21, 22 |
| Gaming | 23 |
| Reading/learning/production | 13, 14, 15, 24, 25, 45, 46, 47 |
| Other waking activities | All other waking slots |

*Note.* Other waking activities include media activities that were not kept as separate parts (Codes 12, 26–28, 30–36, 39, and 40). Secondary activities recorded in the same slot are not included. SNS = social networking service.

**Table A2**

*Sequential Binary Partition of the Nine-Part Time-Use Composition*

| Coordinate | Contrast | Sleep | Broadcast | Video | Audio | Comm. | Search | Gaming | Reading | Other |
|---|---|---|---|---|---|---|---|---|---|---|
| $z_1$ | Sleep vs. waking time | 1 | −1 | −1 | −1 | −1 | −1 | −1 | −1 | −1 |
| $z_2$ | Seven media parts vs. other waking activities | 0 | 1 | 1 | 1 | 1 | 1 | 1 | 1 | −1 |
| $z_3$ | Receptive vs. interactive/productive media | 0 | 1 | 1 | 1 | −1 | −1 | −1 | −1 | 0 |
| $z_4$ | Broadcast vs. online viewing and listening | 0 | 1 | −1 | −1 | 0 | 0 | 0 | 0 | 0 |
| $z_5$ | Video vs. audio | 0 | 0 | 1 | −1 | 0 | 0 | 0 | 0 | 0 |
| $z_6$ | Connective vs. immersive activities | 0 | 0 | 0 | 0 | 1 | 1 | −1 | −1 | 0 |
| $z_7$ | Communication vs. search/SNS/commerce | 0 | 0 | 0 | 0 | 1 | −1 | 0 | 0 | 0 |
| $z_8$ | Gaming vs. reading/learning/production | 0 | 0 | 0 | 0 | 0 | 0 | 1 | −1 | 0 |

*Note.* Broadcast = broadcast viewing; Video = video/personal broadcasting; Comm. = communication; Search = search/SNS/commerce (SNS = social networking service); Reading = reading/learning/production; Other = other waking activities. For a coordinate with *r* parts coded 1 (geometric mean $g_+$) and *s* parts coded −1 (geometric mean $g_-$), $z = \sqrt{rs/(r+s)}\,\ln(g_+/g_-)$. Parts coded 0 do not enter the coordinate.

**Zero-Replacement Equations**

For respondent $i$, let $x_{ij}$ be the number of slots recorded for part $j$, so that the row total is $n_i = 288$. Geometric Bayesian-multiplicative (GBM) replacement (Martín-Fernández et al., 2015) sets the prior location $t_{ij}$ from the part totals of all other respondents and computes the share $r_{ij}$ assigned to a zero cell as follows:

$$t_{ij} = \frac{\sum_{\ell \neq i} x_{\ell j}}{\sum_{h=1}^{9} \sum_{\ell \neq i} x_{\ell h}}, s_i = \left( \prod_{j=1}^{9} t_{ij} \right)^{-1/9}, r_{ij} = t_{ij} \frac{s_i}{n_i + s_i}.$$

If $r_{ij}$ exceeds the smallest observed nonzero share $m_j$ of part $j$, it is replaced by $0.65\, m_j$ (Palarea-Albaladejo & Martín-Fernández, 2015). After these shares are inserted for the zero parts, the nonzero parts are multiplied by a common factor so that all shares sum to 1, which preserves the ratios among the nonzero parts. Because every row sums to 288, $t_{ij}$ for a zero cell is the same for all respondents with a zero in part $j$; moreover, for a respondent with zeros in two parts, $s_i$ cancels in the ratio of the two replacement values, so this ratio does not depend on the respondent (hence the result for $z_8$ in the Zeros and Point Masses section). Our implementation of GBM replacement is written in Python and included in the analysis package. To reproduce it with `cmultRepl` in the R package zCompositions (Version 1.5.0-3 or later), set `z.delete = FALSE`; otherwise, the gaming part, which has more than 80% zeros, is dropped. Table 1 lists the GBM replacement values and log gaps by part.

Multiplicative replacement (Martín-Fernández et al., 2003) inserts the same value δ (min/day) for every zero and multiplies the nonzero parts of a respondent with $c_i$ zeros by $(1440 - \delta c_i)/1440$. The multiplicative-replacement condition in Table B1 uses the conventional value δ = 3.25 min/day, which treats the minimum recording unit (5 min/day) as the detection limit and inserts 65% of it (Martín-Fernández et al., 2003); under this condition, every part has a replacement value of 3.25 min/day and a log gap of 0.43. The sensitivity analysis in Appendix B varies δ from 5 to 0.1 min/day.

# Appendix B

## Sensitivity Analyses

### Classification Under Alternative Analysis Conditions

Table B1 compares the classifications from models fitted under alternative analysis conditions with that of the main analysis, the unweighted five-profile common-covariance model.

**Table B1**

*Classification Comparisons Across Analysis Conditions and Gaming Recording Status*

| Condition | Comparison | Identical classification, % | ARI | Classified differently, *n* | Remarks |
|---|---|---|---|---|---|
| Survey weights and household clustering | Main analysis (unweighted, $K = 5$) | 99.8 | .994 | 18 | Same structure of three gaming recorder profiles and two nonrecorder profiles |
| Multiplicative replacement ($\delta = 3.25$ min/day) | Main analysis | 99.2 | .980 | 65 | Three gaming recorder profiles retained; 35 recorders (3.9%, all with 5 min/day or less) classified into mixed profiles |
| Local independence (Mplus default) | SBP basis vs. Helmert basis ($K = 5$, unweighted) | 62.1 | .435 | 3,066 | With covariances among indicators fixed at 0, the classification depends on how the coordinates are defined (SBP) |
| Profile-specific covariance model, $K = 2$ | Numbers of profiles differ, so only the composition of profiles by gaming recording status is compared | | | | BIC 144,867.8; no recorder profile, one nonrecorder profile, and one mixed profile; gaming-related variance 0.17 in the nonrecorder profile vs. 4.6 in the other profile |
| Profile-specific covariance model, $K = 3$ | Numbers of profiles differ, so only the composition of profiles by gaming recording status is compared | | | | BIC 141,528.2; no recorder profile, one nonrecorder profile, and two mixed profiles; gaming-related variance 0.17 in the nonrecorder profile vs. 4.1–4.4 in the other profiles |

*Note.* Identical classification and the adjusted Rand index (ARI; Hubert & Arabie, 1985) were computed after matching profile labels across solutions. SBP = sequential binary partition; BIC = Bayesian information criterion. The survey-weighted model was estimated with TYPE = COMPLEX MIXTURE, and the multiplicative-replacement row is the $\delta = 3.25$ min/day, $K = 5$ solution in Table B3. The local-independence row compares the classifications obtained with the SBP basis and with the Helmert basis under the Mplus default, which fixes the covariances among indicators at 0. The profile-specific covariance model is an unweighted model with a separate covariance matrix for each profile. Because the gaming-related variance of its

nonrecorder profile (the within-profile variance along the centered log-ratio [clr] axis of gaming in ilr space) was only about 1/25 to 1/27 of that of the other profiles, its BIC was much smaller than that of the main analysis ($K$ = 2: 169,393.4; $K$ = 3: 163,130.6). Results for $K$ = 4, whose best log-likelihood was not replicated, and the eigenvalues and condition numbers of the covariance matrices are reported in Section 6 of the technical notes in the analysis package. Blank cells are not applicable.

## Varying the Zero-Replacement Value

Zeros in the nine parts were replaced by multiplicative replacement with δ = 5, 4, 3.25, 2, 1.5, 1, 0.5, 0.25, and 0.1 min/day, and ilr coordinates were built with the same SBP. For each condition, the unweighted common-covariance models with $K$ = 2 and $K$ = 5 were then estimated. As a reference, the GBM condition (the coordinates of the main analysis) was estimated with the same procedure. All models were estimated in Mplus 9.1 with the same starting-value settings as the main analysis. Because the point mass made the best log-likelihood of the $K$ = 5 model hard to replicate with random starts alone, we also ran each model with a preliminary solution as starting values and kept the solution with the larger log-likelihood (Section 4 of the technical notes in the analysis package describes the procedure and reports the replications by condition). All values in Tables B2–B4 and Figure 2 come from Mplus solutions. Under every condition, the $K$ = 2 solution was replicated 201–433 times with random starts alone. The $K$ = 5 solutions were replicated 0–12 times with random starts alone and 22–206 times in the runs that used preliminary solutions as starting values; the Results section therefore describes the $K$ = 5 solutions mainly in terms of their profile structure. The solution for the conventional value (δ = 3.25 min/day) reported in the Results section, however, was reached with the same best log-likelihood by both kinds of runs (twice with random starts and 118 times with preliminary-solution starting values). Table B2 summarizes the two-profile solutions, and Tables B3 and B4 describe the structure and composition of the five-profile solutions.

**Table B2**

*Two-Profile Solutions and Gaming Recording Status by Zero-Replacement Value*

| δ, min/day | Log gap | Log-likelihood | ARI with gaming recording status | Best-matching part (ARI) | Recorders in the mixed profile, *n* (%) | Profile sizes |
|---|---|---|---|---|---|---|
| 5.00 | 0.00 | −74,676.5 | .830 | Gaming (.830) | 206 (23.0) | 7,400/689 |
| 4.00 | 0.22 | −77,894.9 | .849 | Gaming (.849) | 184 (20.6) | 7,378/711 |
| 3.25 | 0.43 | −80,785.4 | .882 | Gaming (.882) | 145 (16.2) | 7,339/750 |
| 2.00 | 0.92 | −87,152.2 | .946 | Gaming (.946) | 68 (7.6) | 7,262/827 |
| 1.50 | 1.20 | −90,477.0 | .972 | Gaming (.972) | 36 (4.0) | 7,230/859 |
| 1.00 | 1.61 | −94,887.0 | .972 | Gaming (.972) | 35 (3.9) | 7,229/860 |
| 0.50 | 2.30 | −101,808.8 | .996 | Gaming (.996) | 5 (0.6) | 7,199/890 |
| GBM | 2.46 | −84,458.2 | 1.000 | Gaming (1.000) | 0 (0.0) | 7,194/895 |
| 0.25 | 3.00 | −107,799.9 | 1.000 | Gaming (1.000) | 0 (0.0) | 7,194/895 |
| 0.10 | 3.91 | −114,888.6 | 1.000 | Gaming (1.000) | 0 (0.0) | 7,194/895 |

*Note.* Log gap = ln(5/δ). The GBM row uses the coordinates of the main analysis; its log gap is that of gaming (replacement value 0.43 min/day; 1.58–4.49 min/day for the other parts; Table 1). The best-matching part is the media activity whose recording status has the largest ARI with the two-profile classification. Percentages in parentheses are relative to the 895 gaming recorders. Profile sizes are listed in descending order. All solutions are unweighted Mplus solutions. ARI = adjusted Rand index; GBM = geometric Bayesian-multiplicative.

**Table B3**

*Structure of the Five-Profile Solutions by Zero-Replacement Value*

| δ, min/day | Log gap | Log-likelihood | Recorder / nonrecorder / mixed profiles | Recorders in recorder profiles, % | Nonrecorders in nonrecorder profiles, % | ARI with the main solution | Profile sizes |
|---|---|---|---|---|---|---|---|
| 5.00 | 0.00 | −69,023.2 | 3/0/2 | 86.9 | 0.0 | .317 | 4,980/2,331/320/283/175 |
| 4.00 | 0.22 | −71,947.7 | 3/0/2 | 95.8 | 0.0 | .974 | 6,778/454/349/312/196 |
| 3.25 | 0.43 | −74,483.9 | 3/0/2 | 96.1 | 0.0 | .980 | 6,786/443/349/314/197 |
| 2.00 | 0.92 | −80,336.1 | 3/0/2 | 96.3 | 0.0 | .973 | 6,802/425/357/309/196 |
| 1.50 | 1.20 | −83,615.3 | 3/1/1 | 98.2 | 5.8 | .977 | 6,794/416/365/317/197 |
| 1.00 | 1.61 | −87,687.5 | 3/2/0 | 100.0 | 100.0 | .983 | 6,784/410/372/324/199 |
| 0.50 | 2.30 | −93,847.1 | 3/2/0 | 100.0 | 100.0 | .976 | 6,786/408/368/365/162 |
| GBM | 2.46 | −77,532.8 | 3/2/0 | 100.0 | 100.0 | 1.000 | 6,753/441/369/325/201 |
| 0.25 | 3.00 | −98,735.1 | 2/3/0 | 100.0 | 100.0 | .383 | 4,266/2,525/721/403/174 |
| 0.10 | 3.91 | −104,349.1 | 2/3/0 | 100.0 | 100.0 | .382 | 4,268/2,525/721/401/174 |

*Note.* Because a five-profile solution has more profiles than the binary recording status has categories, the magnitude of its ARI with recording status is not interpretable. The solutions are therefore described by the numbers of recorder, nonrecorder, and mixed profiles and by the percentages of recorders classified into recorder profiles and of nonrecorders classified into nonrecorder profiles. The GBM row is the main analysis (gaming replacement value 0.43 min/day, log gap 2.46). ARI with the main solution is the ARI with the five-profile solution under the GBM condition, that is, the model reported in Tables 2 and 3. Profile sizes are listed in descending order. ARI = adjusted Rand index; GBM = geometric Bayesian-multiplicative.

**Table B4**

*Composition of the Five-Profile Solutions by Zero-Replacement Value*

| δ, min/day | Profile | *n* | Gaming recorders, *n* (%) | *Mdn* gaming time, min/day | Gaming | Broadcast | Video | Audio | Comm. | Search | Reading |
|---|---|---|---|---|---|---|---|---|---|---|---|
| 5.00 | 1 | 4,980 | 80 (1.6) | 10 | 98.4 | 3.0 | 40.7 | 66.2 | 4.1 | 58.1 | 84.0 |
| 5.00 | 2 | 2,331 | 37 (1.6) | 10 | 98.4 | 11.1 | 24.2 | 42.6 | 6.0 | 31.9 | 0.0 |
| 5.00 | 3 | 320 | 320 (100.0) | 45 | 0.0 | 19.7 | 15.3 | 32.5 | 3.1 | 25.3 | 37.2 |
| 5.00 | 4 | 283 | 283 (100.0) | 20 | 0.0 | 8.8 | 13.8 | 38.2 | 2.8 | 23.3 | 31.4 |
| 5.00 | 5 | 175 | 175 (100.0) | 130 | 0.0 | 45.7 | 14.3 | 45.1 | 15.4 | 40.6 | 45.7 |
| 3.25 | 1 | 6,786 | 33 (0.5) | 5 | 99.5 | 0.0 | 37.2 | 60.1 | 4.2 | 50.9 | 59.4 |
| 3.25 | 2 | 443 | 2 (0.5) | 5 | 99.5 | 90.7 | 13.1 | 41.8 | 13.5 | 35.7 | 29.8 |
| 3.25 | 3 | 349 | 349 (100.0) | 40 | 0.0 | 14.6 | 14.0 | 33.5 | 3.4 | 26.1 | 36.4 |
| 3.25 | 4 | 314 | 314 (100.0) | 15 | 0.0 | 7.0 | 14.3 | 37.6 | 1.9 | 21.3 | 34.1 |
| 3.25 | 5 | 197 | 197 (100.0) | 120 | 0.0 | 50.8 | 14.7 | 43.1 | 14.2 | 40.6 | 38.6 |
| 0.50 | 1 | 6,786 | 0 (0.0) | | 100.0 | 0.0 | 37.2 | 60.0 | 4.2 | 50.9 | 59.3 |
| 0.50 | 2 | 408 | 0 (0.0) | | 100.0 | 98.3 | 12.5 | 42.2 | 13.7 | 37.0 | 29.7 |
| 0.50 | 3 | 368 | 368 (100.0) | 55 | 0.0 | 0.0 | 17.1 | 34.8 | 5.4 | 27.7 | 38.6 |
| 0.50 | 4 | 365 | 365 (100.0) | 15 | 0.0 | 3.3 | 15.1 | 38.4 | 2.2 | 24.1 | 35.9 |
| 0.50 | 5 | 162 | 162 (100.0) | 80 | 0.0 | 100.0 | 8.0 | 42.0 | 12.3 | 37.0 | 32.7 |
| GBM | 1 | 6,753 | 0 (0.0) | | 100.0 | 0.0 | 37.3 | 60.1 | 4.1 | 51.0 | 59.4 |
| GBM | 2 | 441 | 0 (0.0) | | 100.0 | 90.9 | 12.9 | 42.0 | 13.6 | 36.1 | 29.9 |
| GBM | 3 | 369 | 369 (100.0) | 40 | 0.0 | 15.7 | 14.1 | 33.9 | 3.8 | 24.7 | 35.0 |
| GBM | 4 | 325 | 325 (100.0) | 15 | 0.0 | 4.0 | 15.4 | 38.5 | 1.8 | 23.7 | 36.3 |
| GBM | 5 | 201 | 201 (100.0) | 120 | 0.0 | 51.2 | 14.4 | 42.8 | 13.9 | 40.8 | 39.3 |
| 0.25 | 1 | 4,266 | 0 (0.0) | | 100.0 | 0.0 | 0.0 | 52.9 | 3.1 | 39.5 | 51.4 |
| 0.25 | 2 | 2,525 | 0 (0.0) | | 100.0 | 0.0 | 100.0 | 72.0 | 6.0 | 70.0 | 72.7 |

| δ, min/day | Profile | *n* | Gaming recorders, *n* (%) | *Mdn* gaming time, min/day | Gaming | Broadcast | Video | Audio | Comm. | Search | Reading |
|---|---|---|---|---|---|---|---|---|---|---|---|
| 0.25 | 3 | 721 | 721 (100.0) | 30 | 0.0 | 0.0 | 16.4 | 36.6 | 3.9 | 26.1 | 37.3 |
| 0.25 | 4 | 403 | 0 (0.0) | | 100.0 | 99.5 | 12.4 | 42.4 | 13.9 | 37.2 | 29.3 |
| 0.25 | 5 | 174 | 174 (100.0) | 75 | 0.0 | 100.0 | 7.5 | 41.4 | 11.5 | 35.6 | 32.8 |

*Note.* GBM rows are the main analysis (gaming replacement value 0.43 min/day). Profiles are numbered in descending order of size and do not correspond to P1–P5 in the main text. Gaming recorders, *n* (%), gives the number and percentage of the profile's members who recorded gaming, and *Mdn* gaming time is their median daily gaming time. The columns from Gaming to Reading give the percentage of the profile's members who never recorded the activity in the 3 days. Broadcast = broadcast viewing; Video = video/personal broadcasting; Comm. = communication; Search = search/SNS/commerce; Reading = reading/learning/production. At δ = 5 min/day, the nonrecorders split by whether they recorded reading/learning/production, and 117 recorders with an average of 10 min/day or less (13.1% of all recorders) were classified into these two profiles (Profiles 1 and 2). At δ = 3.25 min/day, the solution consists, as in the main analysis (GBM), of recorder profiles at three levels of gaming time (medians of 15, 40, and 120 min/day) and two profiles made up mostly of nonrecorders; δ = 4, 2, 1.5, and 1 min/day yield the same structure (Table B3). At δ = 0.5 min/day, the gaming recorders also split by whether they recorded broadcast viewing, and the 162 recorders who did not record broadcast viewing formed a separate profile; at δ = 0.25 min/day, the five profiles were defined by combinations of the recording statuses of gaming, broadcast viewing, and video/personal broadcasting. The composition at δ = 0.1 min/day is nearly identical to that at δ = 0.25 min/day (see the profile sizes in Table B3). Blank cells are not applicable because the profile has no gaming recorders. GBM = geometric Bayesian-multiplicative; SNS = social networking service.

# Appendix C

## Model Estimation, Number of Profiles, and Gaming Recording Status

### Estimation and Replication

All models were estimated by robust maximum likelihood (MLR) in Mplus 9.1 with 3,000 random starts and 750 final-stage optimizations. The adjusted Lo–Mendell–Rubin likelihood ratio test (aLMR; TECH11) was computed for $K = 2–8$. Because of the computational burden, the parametric bootstrap likelihood ratio test (BLRT; TECH14) was computed only for $K = 2–6$, with 200 bootstrap draws (LRTBOOTSTRAP = 200, LRTSTARTS = 0 0 500 200; Asparouhov & Muthén, 2012). Of the 750 final-stage optimizations, the best log-likelihood was replicated 433, 55, 6, 6, 36, 5, and 3 times for $K = 2–8$, respectively. The log-likelihoods of the null $(K - 1)$-profile models reestimated within TECH11 ($K = 2–8$) and TECH14 ($K = 2–6$) all equaled the best log-likelihoods of the corresponding $(K - 1)$-profile models (Asparouhov & Muthén, 2012). When the selected $K = 5$ model was reestimated with twice as many random starts, the same best log-likelihood was replicated 14 times. Figure C1 shows the information criteria and the decrease in the Bayesian information criterion (BIC).

**Figure C1**

*Information Criteria and Decrease in BIC by Number of Profiles*

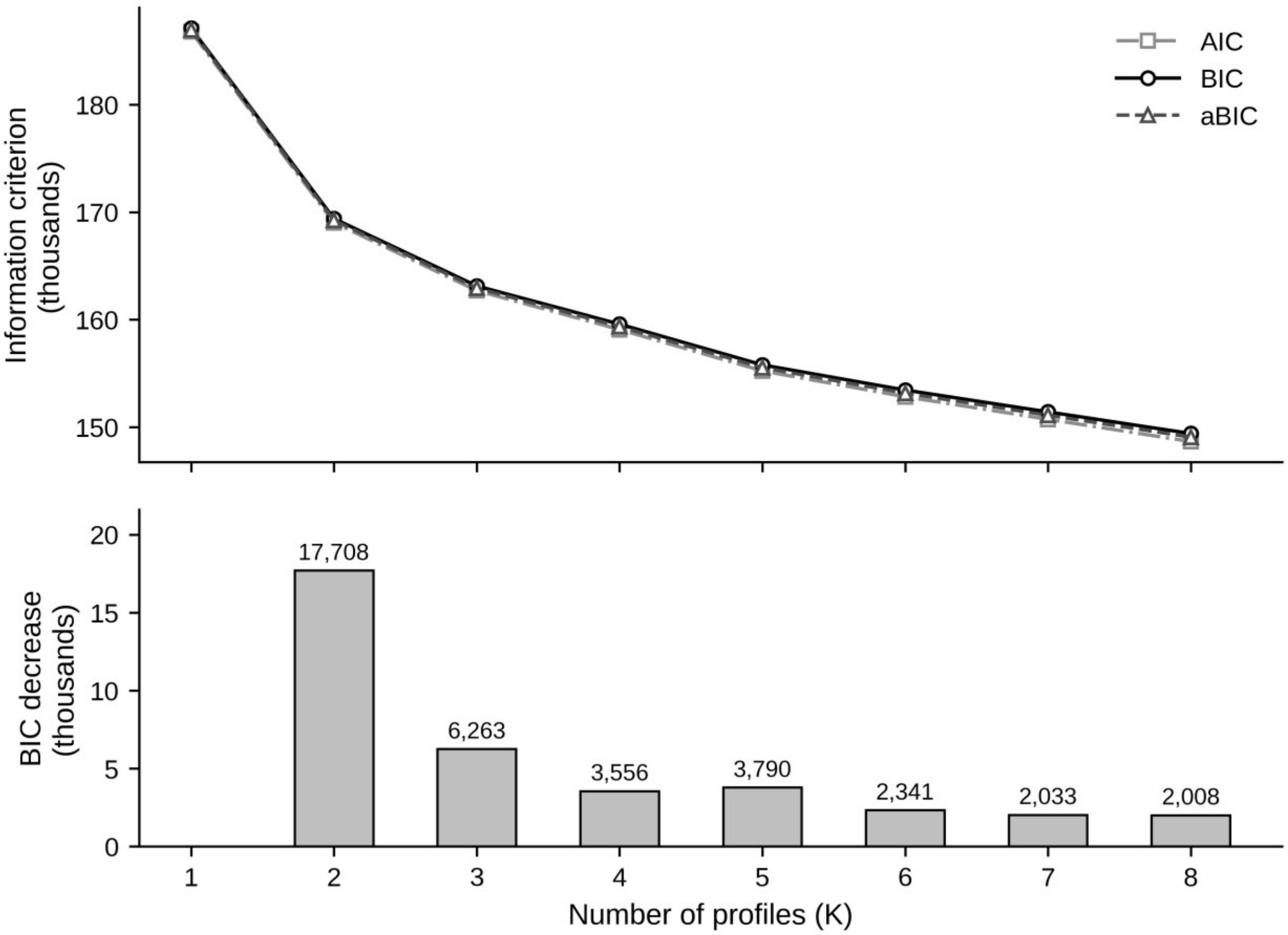

*Note.* The top panel shows AIC, BIC, and aBIC of the unweighted common-covariance models (GBM replacement), and the bottom panel shows the decrease in BIC from the model with one fewer profile, both in thousands. The decrease in BIC was 17,708 from $K$ = 1 to 2, 6,263 from $K$ = 2 to 3, 3,556 from $K$ = 3 to 4, and 3,790 from $K$ = 4 to 5, and it fell to 2,341 from $K$ = 5 to 6, 2,033 from $K$ = 6 to 7, and 2,008 from $K$ = 7 to 8. AIC = Akaike information criterion; BIC = Bayesian information criterion; aBIC = sample-size-adjusted BIC; GBM = geometric Bayesian-multiplicative.

**Profiles by Gaming Recording Status**

Table C1 counts, for each model with $K = 2–8$, the recorder profiles (consisting only of gaming recorders), the nonrecorder profiles (consisting only of nonrecorders), and the mixed profiles (containing both). Individual classifications were verified by matching the SAVEDATA output of the same run with the respondent identifiers. No model had a mixed profile.

**Table C1**

*Number of Profiles by Gaming Recording Status*

| *K* | Recorder profiles | Nonrecorder profiles | Mixed profiles |
|---|---|---|---|
| 2 | 1 | 1 | 0 |
| 3 | 2 | 1 | 0 |
| 4 | 3 | 1 | 0 |
| 5 | 3 | 2 | 0 |
| 6 | 3 | 3 | 0 |
| 7 | 4 | 3 | 0 |
| 8 | 4 | 4 | 0 |

*Note.* Each row is a separately estimated unweighted common-covariance model. The one-profile model is omitted because all respondents belong to one profile by definition. At $K = 6$, 7, and 8, a separate profile of 42, 45, and 43 nonrecorders, respectively (0.5%–0.6% of the sample), was formed. At $K = 7$, the recorders split into four profiles; at $K = 8$, the respondents who recorded broadcast viewing but not gaming split into a group of 1,968, all of whom recorded reading/learning/production, and a group of 4,758, of whom 84.0% did not.

## Appendix D

## Profile Characteristics

Table D1 gives the survey-weighted classification shares of the five profiles with 95% confidence intervals, together with the mean age and percentage of men in each profile. Figure D1 shows the gaming time of the gaming recorders by profile, and Figure D2 shows the mean time spent on each media activity by profile.

**Table D1**

*Survey-Weighted Classification Shares With 95% Confidence Intervals, Age, and Sex by Profile*

| Profile | *n* | Share, % [95% CI] | Mean age, years | Male, % |
|---|---|---|---|---|
| P1 Gaming recorded: long | 201 | 3.0 [2.4, 3.7] | 32.2 | 90.7 |
| P2 Gaming recorded: medium | 369 | 5.8 [4.8, 6.8] | 38.6 | 81.6 |
| P3 Gaming recorded: short | 325 | 5.4 [4.4, 6.4] | 45.6 | 68.4 |
| P4 Gaming and broadcast not recorded | 441 | 6.4 [5.3, 7.4] | 33.9 | 43.0 |
| P5 Gaming not recorded, broadcast recorded | 6,753 | 79.4 [77.6, 81.1] | 54.5 | 44.7 |

*Note.* Survey-weighted descriptive statistics computed from the modal assignments, without accounting for classification error. Sizes (*n*) are unweighted numbers of respondents. Confidence intervals (CIs) for the shares were obtained from Taylor-linearized standard errors that account for household clustering, using a normal approximation. In the whole analysis sample, the survey-weighted mean age is 51.1 years and the survey-weighted proportion of male respondents is 49.4%.

**Figure D1**

*Daily Gaming Time of Gaming Recorders by Profile*

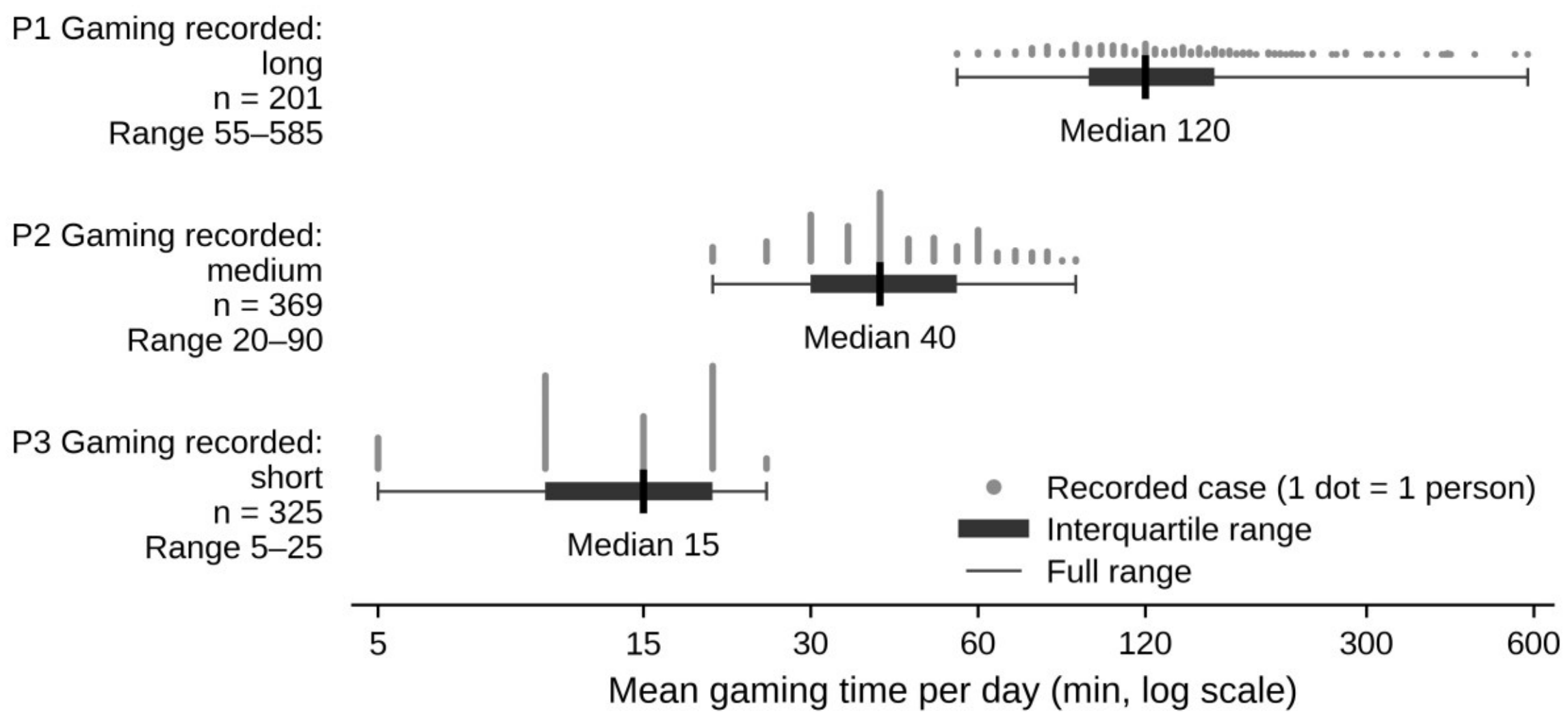

*Note.* $n$ = 895. Unweighted distribution of daily gaming time before replacement. Each dot is one recorder, and dots with the same value are stacked vertically. Thick horizontal bars are interquartile ranges, vertical marks are medians, and thin horizontal lines run from the minimum to the maximum. The horizontal axis is on a log scale. The ranges of P1–P3 are 55–585, 20–90, and 5–25 min/day, and their medians are 120, 40, and 15 min/day; the boundary intervals (20–25 and 55–90 min/day) contain respondents from two adjacent profiles. The median for all recorders is 35 min/day; 22.1% of them recorded 15 min/day or less and 32.1% recorded 60 min/day or more (survey-weighted, 23.5% and 31.1%).

**Figure D2**

*Mean Daily Time Spent on Media Activities by Profile in the Five-Profile Model*

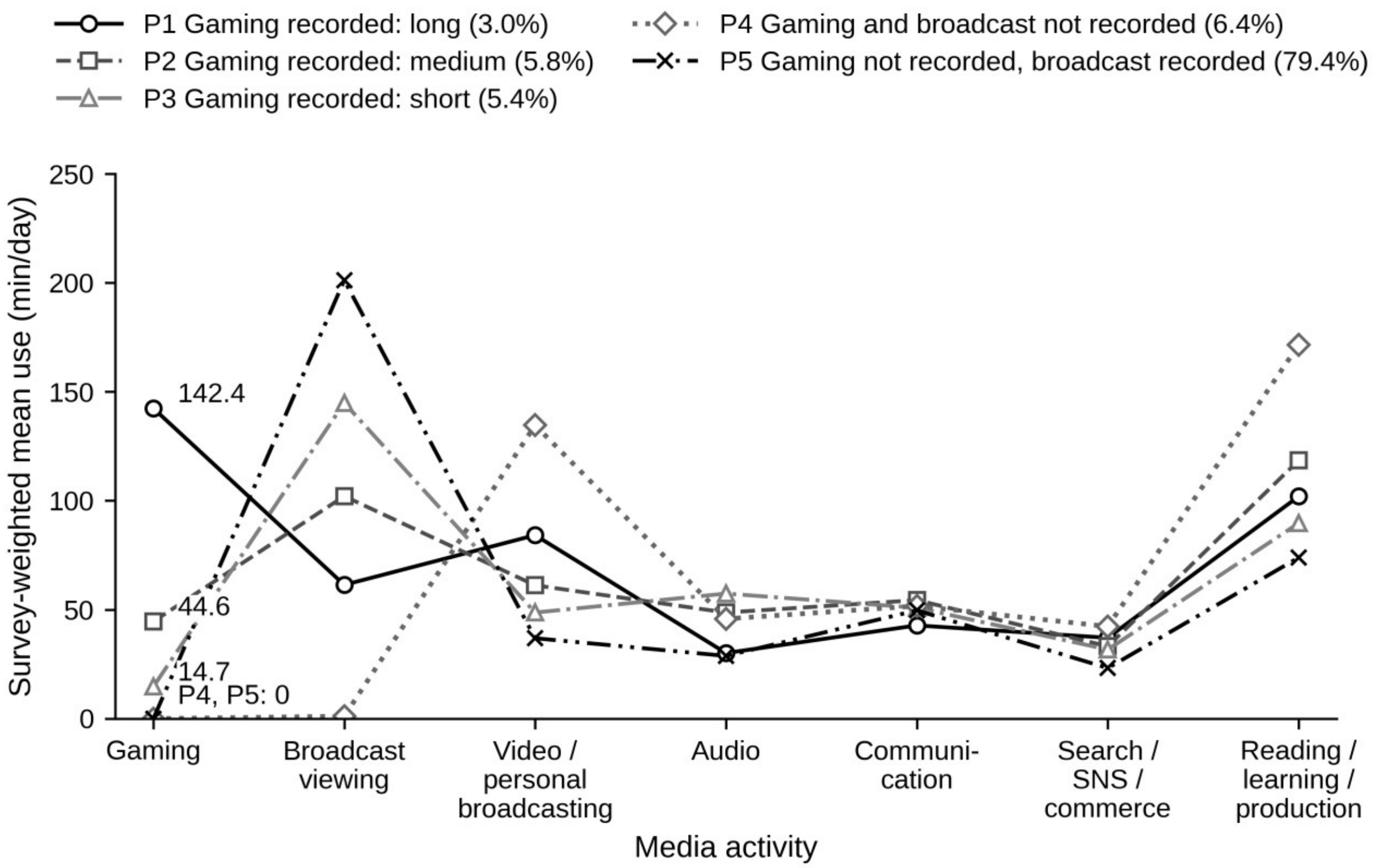

*Note.* Survey-weighted means of primary-activity time before replacement, by modal assignment; these are not back-transformed ilr means. Percentages in the legend are survey-weighted classification shares. Of the five profiles, P4 had the longest mean times for reading/learning/production (171.5 min/day) and video/personal broadcasting (134.7 min/day), and in P5 broadcast viewing (201.2 min/day) accounted for 48.7% of total media time.

ilr = isometric log-ratio; SNS = social networking service.